\documentclass{icrc29}
\usepackage{graphicx,amssymb,amsmath,times}

\begin{document}
\title[The EEE Project]{The EEE Project}

\author[A. Zichichi et al.]{R. Antolini, 
R. Baldini Ferroli, 
M. Caporaloni, 
A. Chiavassa, 
L. Cifarelli, 
F. Cindolo, \newauthor
E. Coccia, 
S. De Pasquale,
M. Garbini, 
C. Gustavino, 
D. Hatzifotiadou, 
G. Imponente, 
\newauthor
H. Menghetti, 
 G. Piragino, 
G. Sartorelli, 
M. Selvi, 
C. Williams, A. Zichichi
         \\
    Museo Storico della Fisica, Centro Studi e Ricerche ``E. Fermi'', Roma, Italy \\
        Universit\`a di Bologna and INFN, Bologna, Italy \\
        Universit\`a di Torino and INFN, Torino,Italy  \\
        Universit\`a di Salerno and INFN, Salerno, Italy \\
        Universit\`a di Catania and INFN, Catania, Italy\\
        INFN-LNGS, Assergi (Aq), Italy\\
        INFN-LNF, Frascati, Italy   
        }
        
\presenter{Presenter: G. Sartorelli (gabriella.sartorelli@bo.infn.it), \  
uki-imponente-G-abs1-he15-oral}


\maketitle

\begin{abstract}

The  new experiment  ``Extreme Energy Events'' (EEE) to detect 
extensive air showers through muon detection is starting in Italy.
The use of particle detectors based on Multigap Resistive Plate 
Chambers (MRPC)
will allow to determine with a  very high accuracy the direction 
of the axis of cosmic ray showers initiated  by primaries of 
ultra-high energy, together  with a high temporal resolution.
The installation of  many of such 'telescopes' in numerous High 
Schools scattered all over the Italian territory  will also allow 
to investigate coincidences  between multiple primaries
 producing distant showers. 
Here we present the experimental apparatus and its tasks.
\end{abstract}

\section{Introduction}

Since the first evidence of the existence of the Cosmic
Rays (CRs), dating back to Victor Hess in 1912, an increasing 
interest grew,  for the phenomenon of particles
coming from the Space as well as for the detection 
techniques devoted to measure the secondaries produced in 
the interaction with the upper atmosphere and arriving down to the ground.
The higher the energy of an incoming cosmic ray, the 
larger the area to be covered in order to measure 
the induced Extensive Air Shower (EAS) generated so far. 
Given the peculiarities of the CR flux, the interest to 
take many data with various kind of detectors  widely distributed, 
aims to confirm (or not) the indications of primaries 
violating the so-called GZK cut-off (see for example \cite{NW00} 
for a review, and references therein). 
A unified framework for explaining the spectrum behavior 
at energies above $10^{15}$ eV is still missing, in particular
for what regards the details of the slope changes as well as for
the existence of a CRs energy upper limit, or not. \\
The EEE Project is designed for being a very large array of 
particle detectors scattered over the whole Italian territory 
and located inside High Schools buildings and INFN Sections 
and Laboratories. This set up has the peculiar characteristic of having, 
once in operation, 
a double grid pace: a small one (given by the average 
separation of Schools within the same city), order  of hundreds meters, 
aimed to measure a single shower; a larger one, order of tens or
hundreds kilometers, in order to search for correlations between 
two (or more) primaries. The detectors array inside the same city 
is sensible to EAS with an energy threshold $\sim 10^{18}$ eV. Using 
the data coming from different cities it would be possible to search for 
the existence of a ``shower of primaries'' formed outside the Earth's 
atmosphere by means of different mechanisms, from astrophysical 
\cite{FMO83, MTW99, McBL81, PSB76, MS98, GZ60, SS99, IS05} to exotic 
ones (for review \cite{BS00}).\\ 
%
%
The review presented in \cite{IS05} gives a new estimate
of the multiple primaries rate provided some years 
ago in \cite{MTW99}, keeping in 
mind the design and the perspectives of the present experiment.

\section{Detector design}

The basic idea of the detector is that it has to be able to 
reconstruct the directions of the incoming muons produced by the 
interaction of a primary with an air nucleus. The muons mainly 
proceed directly to the ground without interacting, hence the 
detection of many of them allows a very accurate 
measure of the primary's direction.
%
%
%
%
%
\begin{figure}[h]
\begin{center}
\resizebox{\hsize}{0.25\vsize}{
\includegraphics*[width=0.2\textwidth,angle=0,clip]{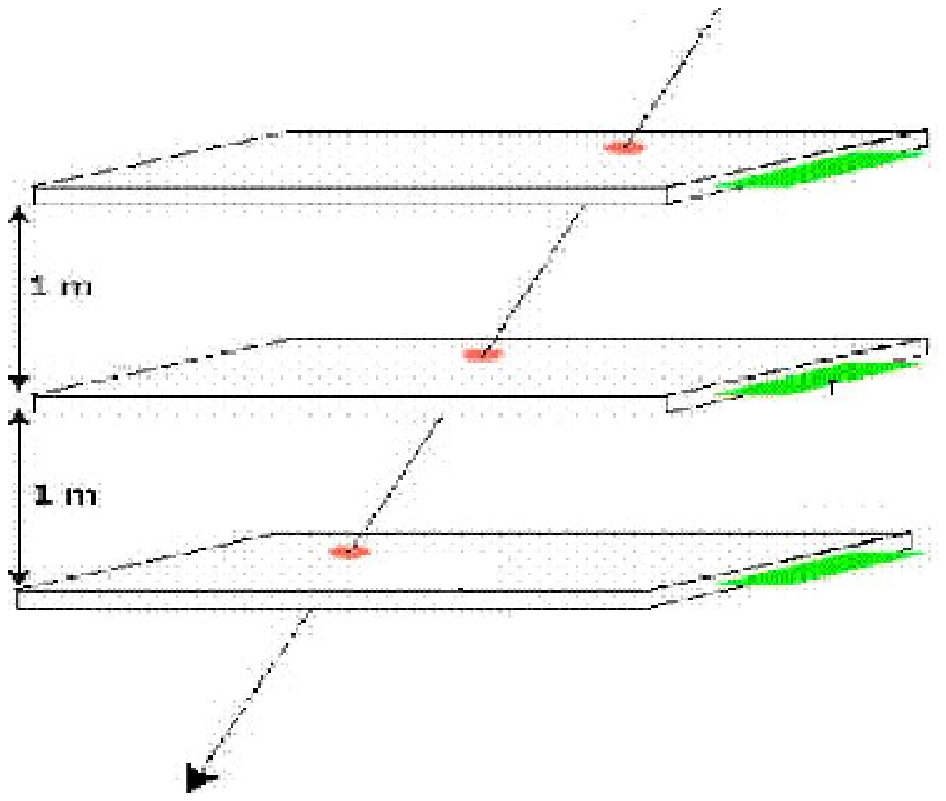}
\includegraphics*[width=0.3\textwidth,angle=0,clip]{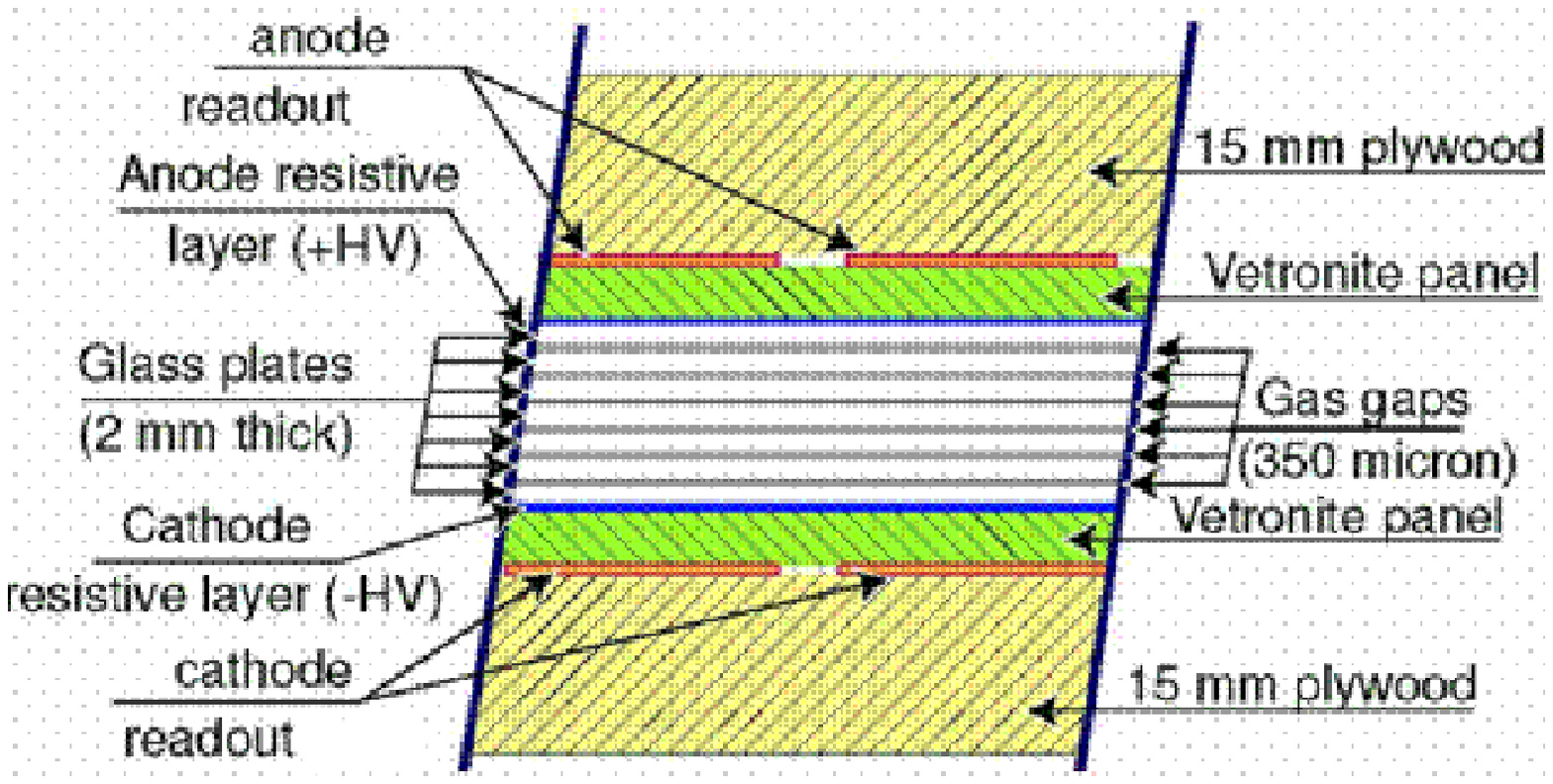}}
\caption{\label{telescope} (left) Scheme of the three MRPC detectors 
constituting the single telescope and (right) prototype of the 
MRPC chamber and inner layers.}
\end{center}
\end{figure}
The detector used in this project 
is a telescope consisting of three Multigap Resistive Plate Chamber 
(MRPC) spaced of $1~m$, one over the other as in as in 
Figure \ref{telescope} (left).
This kind of detector is robust, allows to track the secondary muons 
of an EAS with an angular resolution less than $0.5^{o}$ 
(see Figure \ref{acce} (left)), thanks to a space resolution of the 
track impact point of the order of the centimeter, and time resolutoin 
of the order of hundred of picoseconds.\\
The basic design of the MRPC consists of six gas gaps of $350~\mu m$ 
to enhance streamer-free operation inside two resistive plates of glass 
sheets coated with a resistive paint. Two vetronite panels insulate the 
readout copper strips from the anode and the cathode. The signal initiated by 
a charged particle transversing the detector is formed on such 
strips and proceed to both ends: the time difference of the signals 
arriving at the two strip ends gives the coordinate 
along the copper line. A honeycomb support and aluminum boxes complete the 
mechanics. The two coordinates of the impact point of a muon 
are then measured in each chamber.\\
The gas filling mixture consists of $93\%~C_{2}F_{4}H_{2}$ and 
$7\%~SF_{6}$: the first tests were performed with gas fluxing. Inside the 
Schools the chambers will operate with no gas fluxing due to security 
reasons. The apparatus will be tested to determine the gas refilling 
duty cycle  needed to mantain its very high efficiency 
(close to $100\%$) for many weeks or months.\\





\section{Performance of the telescope and reconstruction of axis direction}

\begin{figure}[h]
\begin{center}
\resizebox{\hsize}{0.3\vsize}{
\includegraphics*[width=4.5cm,angle=0,clip]{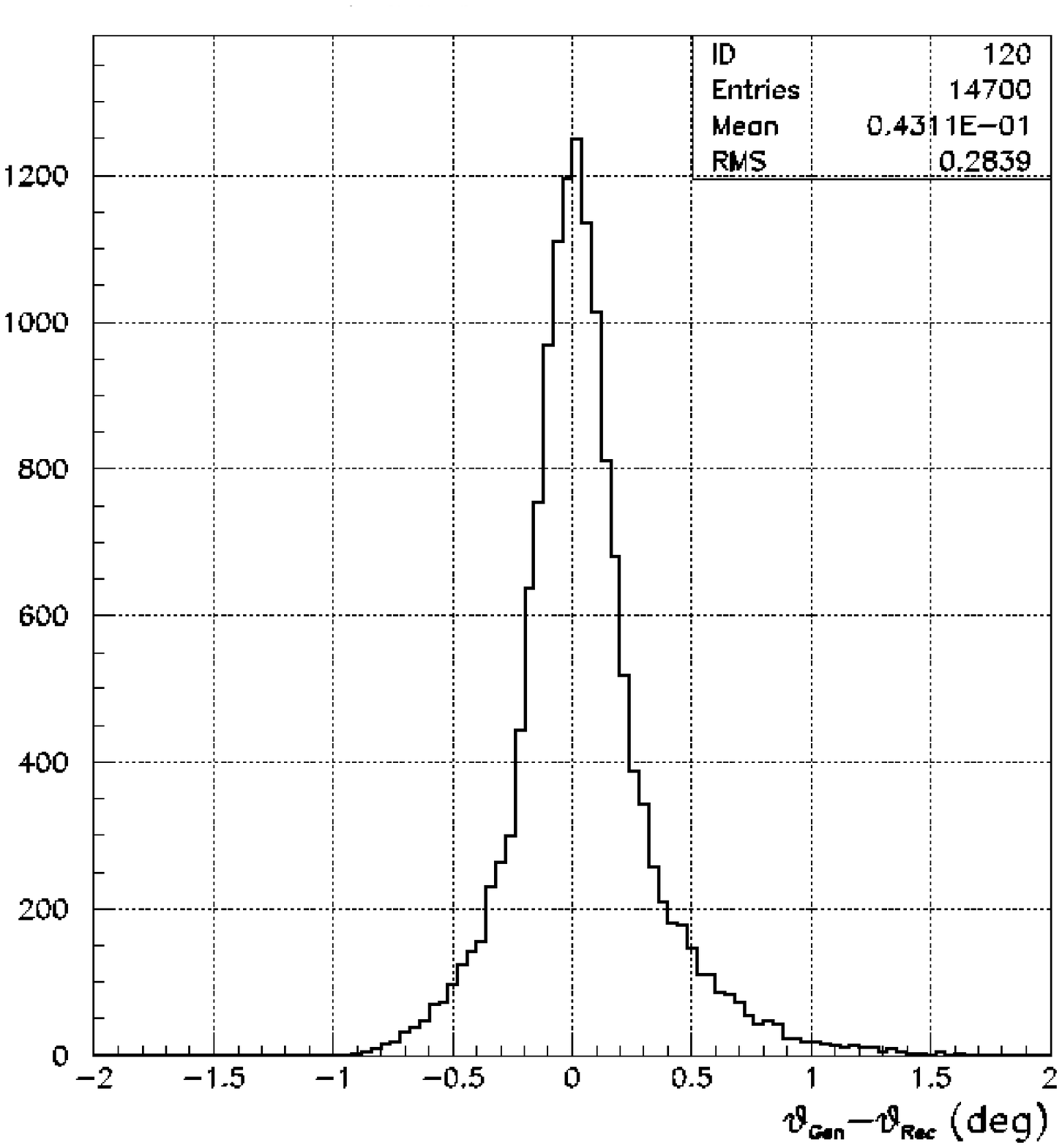}
\includegraphics*[width=5cm,angle=0,clip]{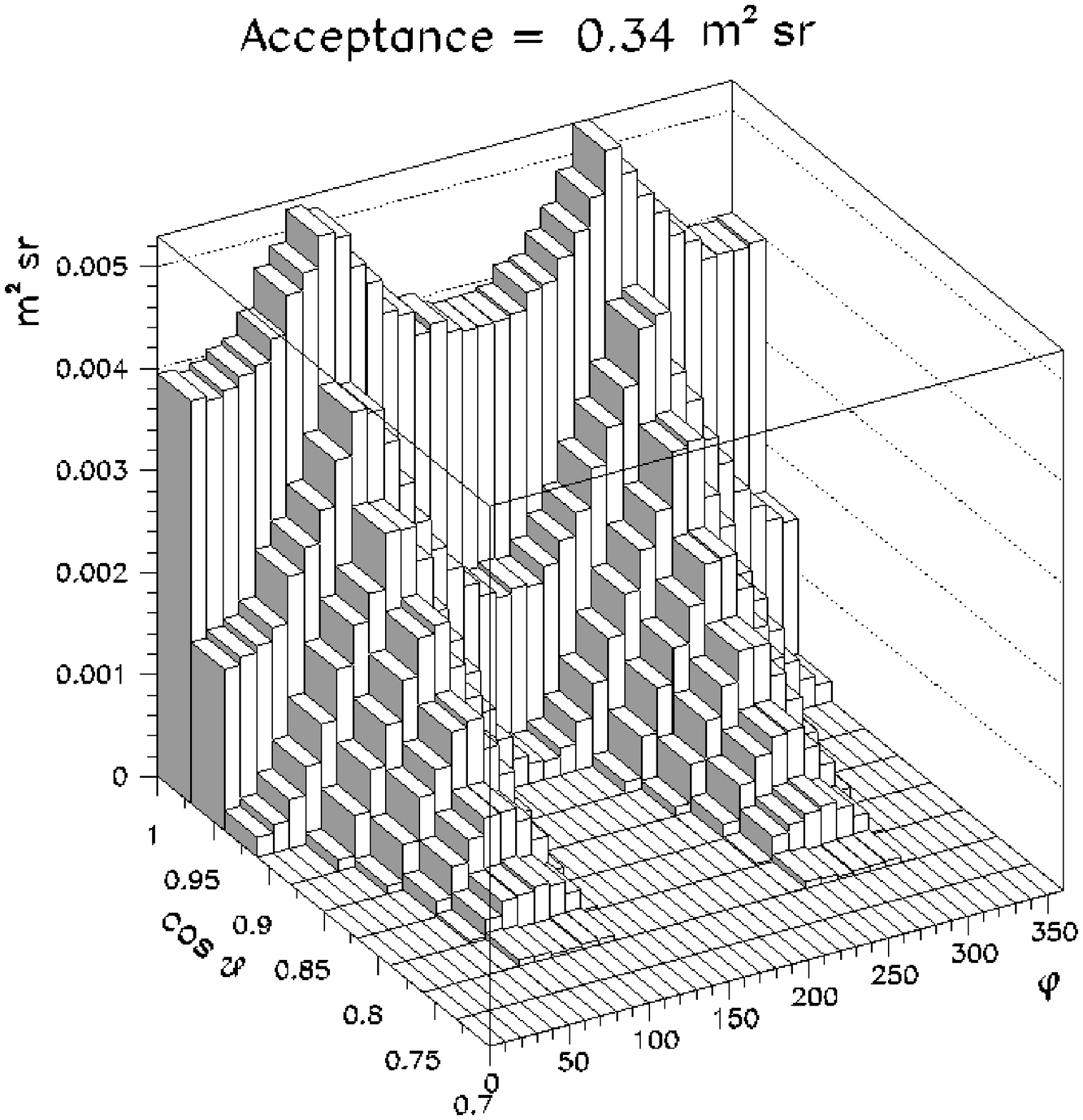}}
\caption{\label{acce} (right) Geometrical acceptance of the 
telescope in bins of $\Phi$ and $\cos(\theta)$. 
The total acceptance is 0.34 $m^2 sr$ and 
(left) difference between the generated and the reconstructed 
muon direction zenith angle.}
\end{center}
\end{figure}

\begin{figure}[h]
\begin{center}
\resizebox{\hsize}{0.3\vsize}{
\includegraphics*[width=4cm,angle=0,clip]{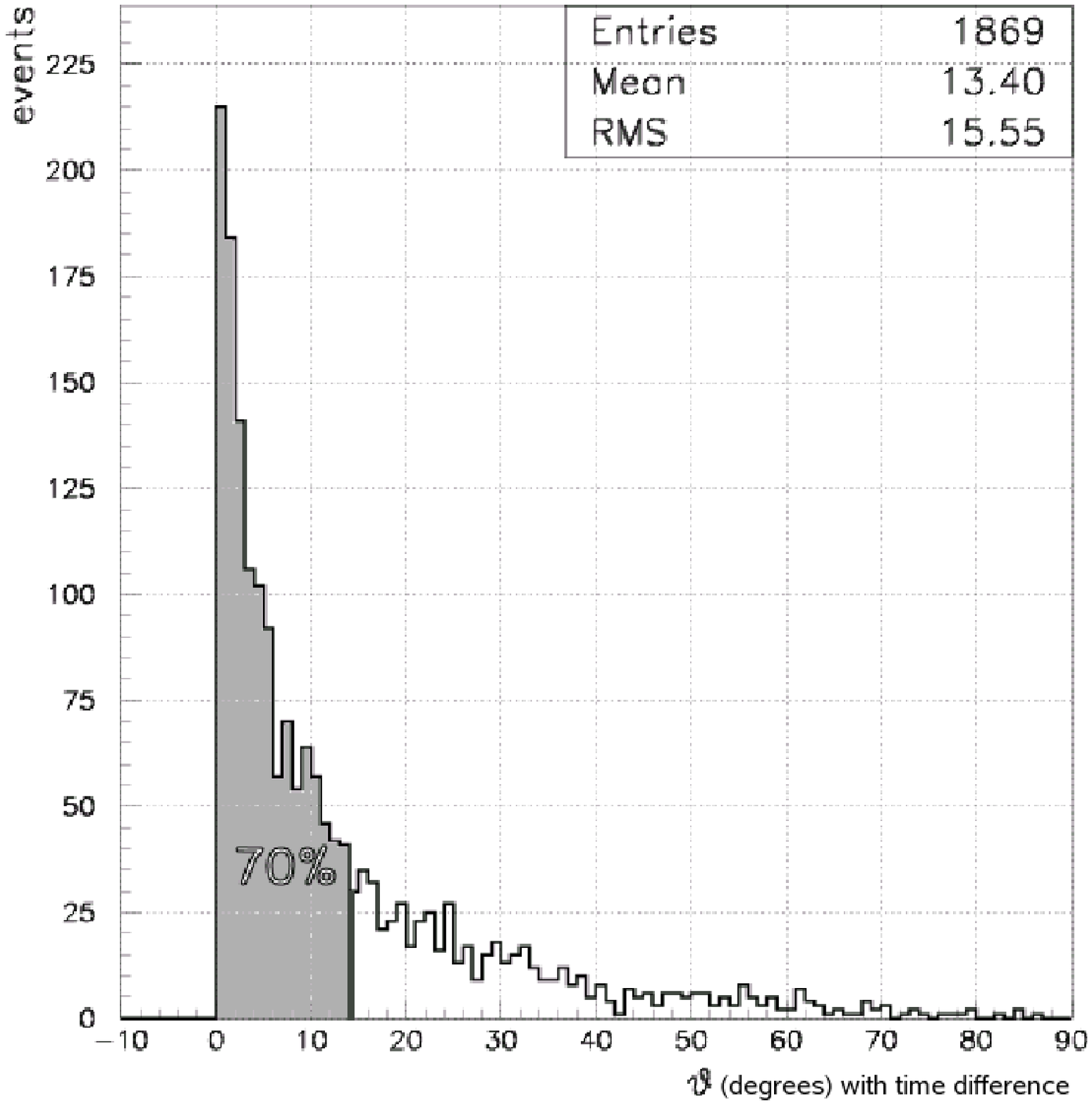}
\includegraphics*[width=4cm,angle=0,clip]{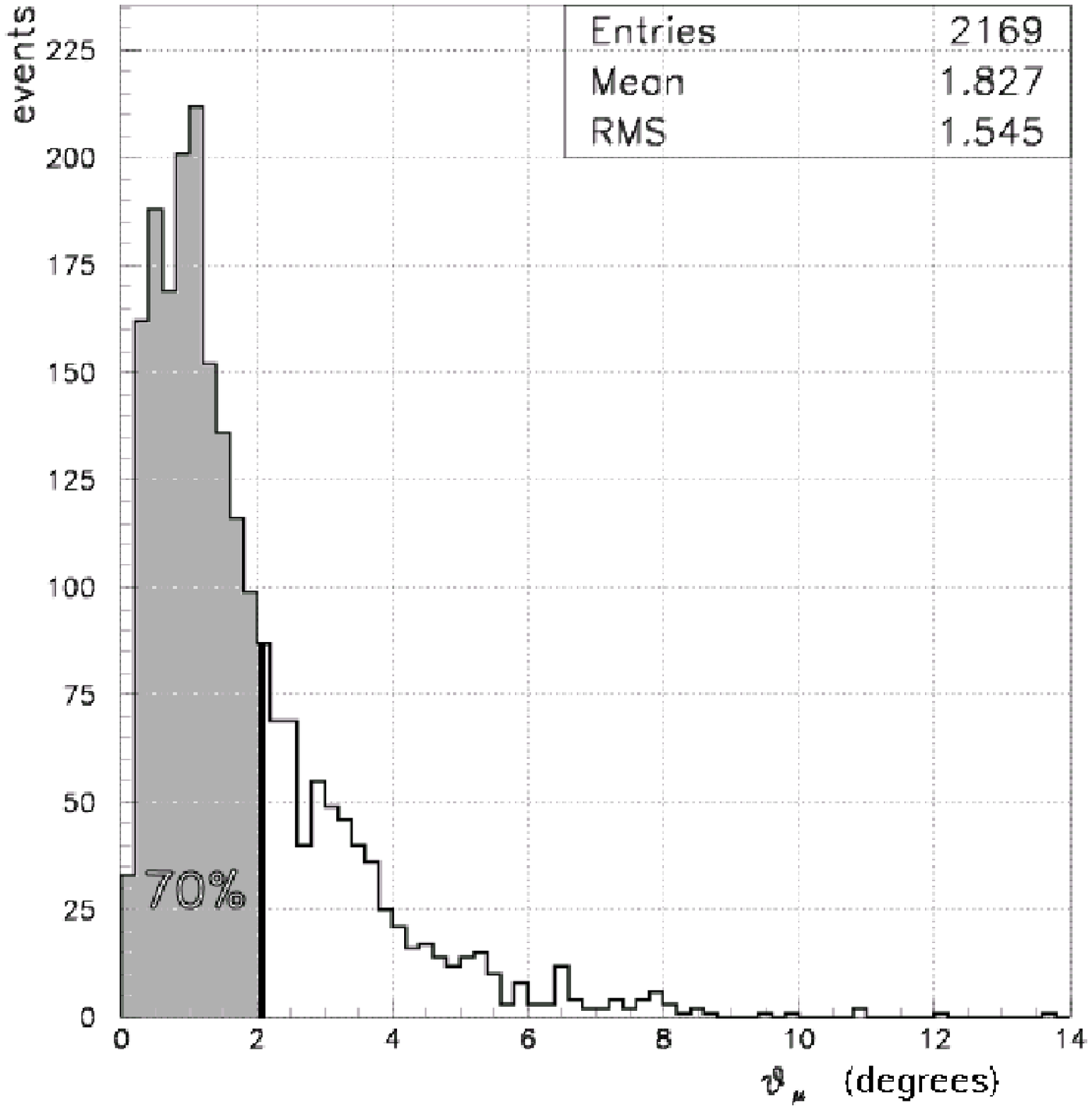}}
\caption{\label{rec} 
Angular resolution of the reconstruction of the shower axis direction
(left) taking into account the arrival time information and 
(right) the recosntructed muons. The shaded areas shows that $70\%$
of the events are reconstructed with an angular uncertainty smaller
than $14^{o}$ (left) and $2^{o}$ (right).}
\end{center}
\end{figure}

In each MRPC chamber of the telescope both the two impact coordinates 
and the crossing time of a muon are measured. From the position of the 
three impact points (one per plane) it is possible to reconstruct the 
direction of the crossing muon.\\
Taking into account the design of the MRPC telescope we performed 
Monte Carlo simulations in order to evaluate the detector geometrical 
acceptance. A sample of $10^7$ muons, with incoming direction uniformly 
sampled on the  upper hemisphere surrounding the detector, has been 
generated. The resulting overall geometrical acceptance, whose details 
are shown in  Figure \ref{acce} (right), is $0.34~m^{2}sr$.  
The telescope acceptance can, of course, be varied 
changing the distance between the MRPC planes so that for 
example, reducing the distance between the three planes, it is 
possible to reconstruct muons at larger zenith angles.\\
Another simulation has been performed in order to evaluate the expected muon 
rate in one telescope, due to the cosmic muons steady flux. The inputs to the simulation 
are the known flux and angular distribution of cosmic muons at ground. With this procedure
we find that the expected  muon rate per telescope is equal to $36~Hz$. That result has 
been confirmed with a full Monte Carlo simulation of the primary flux, the shower production 
and propagation down to the sea level and the muon detection in the telescope.\\
We also studied the telescope capability to reconstruct the shower 
axis direction. Using the position and arrival time of the shower front on 
the detectors at ground (at least three non aligned detectors are needed) 
$70\%$ of the reconstructed shower axes have an angular uncertainty smaller 
than $14^{o}$ (Figure \ref{rec} (left)).\\
Thanks to the telescope excellent tracking capabilities it is possible to 
reconstruct the shower axis direction using the reconstructed muons 
direction. As shown in Figure \ref{rec} (right) $70\%$ of the reconstructed shower 
axes have an angular uncertainty smaller than $2^{o}$. The accuracy 
increases with the number of muons hitting the detectors even if this method can provide 
informations on shower axis direction even when less than three 
telescopes are hit (which is otherwise impossible using arrival times).
%
%
%
%
%

\section{Project plan}

The EEE Project will last at least 10 years and 
has to cover the whole Italian territory 
as much as possible, in order to obtain a grid over more 
than 300.000 km$^2$. 
The actual plan is designed to follow the natural distribution 
of candidate  Schools, characterized by a closer spacing 
within each city of order of hundreds meters  and a larger
one between one city and the other, of order hundreds
of kilometers. Where a INFN Section is present, 
an additional telescope will be installed also there. \\
The financing institutions in Italy  are ``Museo Storico della Fisica and Centro
Studi e Ricerche Enrico Fermi'' in Rome, the MIUR - Ministero dell'Istruzione 
Universit\`a e Ricerca and INFN - Istituto Nazionale di Fisica Nucleare. \\
The present status of the apparatus consists of  the full construction 
and testing of advanced prototypes of  21 MRPC performed in the 
CERN facilities. These first telescopes  will be installed 
in seven piloting Schools by the end of the year 2005, while the construction
of more detectors is ongoing and fastly proceeding. \\


\end{document}